\documentclass[twoside]{article}
\usepackage{fleqn,espcrc1}
\usepackage{amsmath}

\usepackage[dvips]{graphicx}

\newcommand{\etal}{{\it et al.}}

\newcommand{\Tr}{\mbox{Tr}}
\newcommand{\be}{\begin{eqnarray}}
\newcommand{\ee}{\end{eqnarray}}
\newcommand{\bdm}{\begin{displaymath}}
\newcommand{\edm}{\end{displaymath}}
\newcommand{\<}{\langle}
\renewcommand{\>}{\rangle}

\def\spose#1{\hbox to 0pt{#1\hss}}
\def\ltapprox{\mathrel{\spose{\lower 3pt\hbox{$\mathchar"218$}}
 \raise 2.0pt\hbox{$\mathchar"13C$}}}
\def\gtapprox{\mathrel{\spose{\lower 3pt\hbox{$\mathchar"218$}}
 \raise 2.0pt\hbox{$\mathchar"13E$}}}
\def\inapprox{\mathrel{\spose{\lower 3pt\hbox{$\mathchar"218$}}
 \raise 2.0pt\hbox{$\mathchar"232$}}}
\title{Thermodynamics with 2+1 and 3 Flavors of Improved Staggered Quarks
\thanks{Preprint FSU-CSIT-01-50. Talk given by U.M.~Heller at ``Statistical
QCD'', August 26--30, 2001, Bielefeld, Germany.}}
\author{ C.~Bernard
\address{Department of Physics, Washington University, St.~Louis, MO 63130,
USA},
T.~Burch
\address{Department of Physics, University of Arizona, Tucson, AZ 85721, USA}, 
S.~Datta
\address{Department of Physics, Indiana University, Bloomington, IN 47405,
USA},
T.A.~DeGrand
\address{Physics Department, University of Colorado, Boulder, CO 80309, USA},
C.E.~DeTar
\address{Physics Department, University of Utah, Salt Lake City, UT
  84112, USA},
Steven~Gottlieb$\,\null^{\rm c}$,
U.M.~Heller
\address{CSIT, Florida State University, Tallahassee, FL 32306-4120, USA},
K.~Orginos
\address{RIKEN-BNL Research Center,
Brookhaven National Laboratory, Upton, NY 11973-5000},
R.L.~Sugar
\address{Department of Physics, University of California, Santa Barbara,
CA 93106, USA},
and D.~Toussaint$\,\null^{\rm b}$
} %end \author

\begin{document}

\maketitle 

\begin{abstract}
We present preliminary results from exploring the phase diagram of finite
temperature QCD with three degenerate flavors and with two light flavors
and the mass of the third held approximately at the strange quark mass.
We use an order $\alpha_s^2 a^2, a^4$ Symanzik improved gauge action and
an order $\alpha_s a^2, a^4$ improved staggered quark action. The improved
staggered action leads to a dispersion relation with diminished lattice
artifacts, and hence better thermodynamic properties. It decreases the
flavor symmetry breaking of staggered quarks substantially, and we estimate
that at the transition temperature for an $N_t=8$ to $N_t=10$ lattice
{\em all} pions will be lighter than the lightest kaon. Preliminary results
on lattices with $N_t=4$, 6 and 8 are presented.
\end{abstract}

\section{INTRODUCTION}

With the Relativistic Heavy Ion Collider (RHIC) now producing data,
it has become even more important to understand the phase diagram
of QCD at finite temperature, and to determine properties of the
high temperature quark-gluon-plasma phase with confidence, {\it i.e.}
with controlled lattice spacing errors.

It is fairly well established that QCD with two flavors of massless
quarks has a second order finite temperature, chiral symmetry restoring
phase transition. This transition is washed out as soon as the quarks
become massive. QCD with three flavors of massless quarks has a first
order finite temperature, chiral symmetry restoring phase transition,
which is stable for small quark masses. Not well known is
how large the quark masses can be before the phase transition turns second
order and then into a crossover, both for degenerate quarks and especially
for the physically relevant case of two light and one heavier strange quark.

In previous studies, the second question is particularly badly answered
due to the flavor symmetry breaking in Kogut-Susskind quarks, usually used
for this purpose: how can one study the influence of the strange quark
when most of the (non-Goldstone) pions are heavier than the (Goldstone)
kaon?

Adding a few terms to the conventional Kogut-Susskind action, namely
three-link, five-link and seven-link staples and a third-neighbor
coupling, removes all tree-level ${\cal O}(a^2)$ errors
\cite{NAIK,MILC_FATTEST,LEPAGE98}.
This ``Asqtad'' action shows improved flavor
and rotational symmetry \cite{MILC_FATTEST,MILC_Spec}, and, at least
in quenched QCD, good scaling properties \cite{IMP_SCALING}.
This is illustrated in Fig.~\ref{fig:pi05_rho_vs_a}. There, and in other
figures below, we plot results in units of $r_1$, a scale defined in terms
of the static ${\rm Q\bar Q}$ potential by $r_1^2 F_{\rm Q\bar Q static}(r_1)
= 1$, which leads to $r_1 \sim 0.35$ fm \cite{MILC_POT}.

\begin{figure}
\begin{center}
\begin{tabular}{c c}
\includegraphics[width=2.5in]{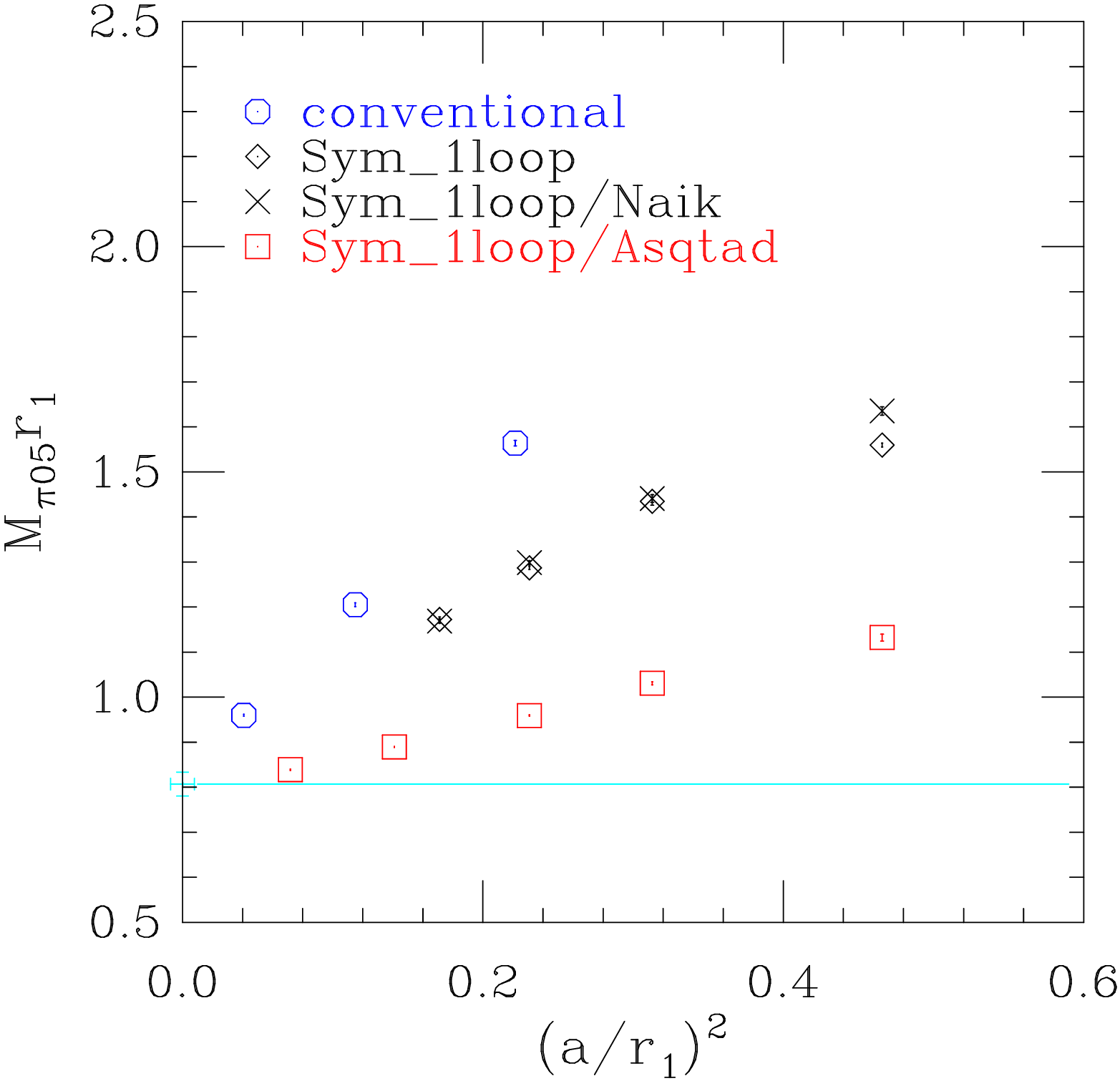}
&
\includegraphics[width=2.5in]{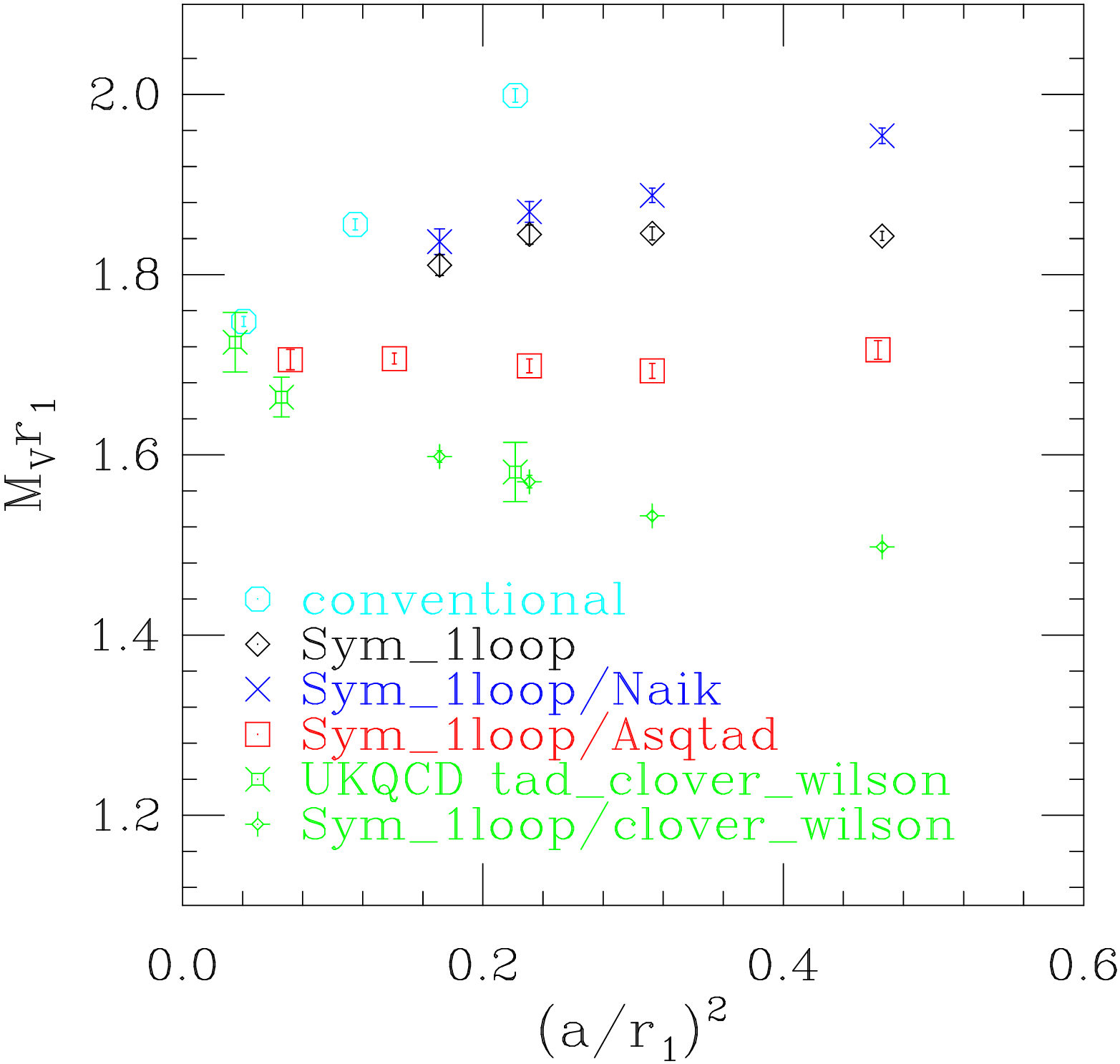}
\end{tabular}
\end{center}
\vspace*{-9mm}
\caption{The mass of the second lightest pion, for fixed lowest pion mass
in units of $r_1$, shown as the horizontal line, as function of the lattice
spacing (left). The ``Asqtad'' action gives the smallest flavor symmetry
breaking. The mass of the vector meson, the rho, as function of the lattice
spacing (right). The ``Asqtad'' action gives the best scaling.
\label{fig:pi05_rho_vs_a} 
}
\vspace*{3mm}
%\vspace*{-10mm}
%\end{figure}
%
%\begin{figure}
\begin{center}
\begin{tabular}{c c}
\includegraphics[width=2.5in]{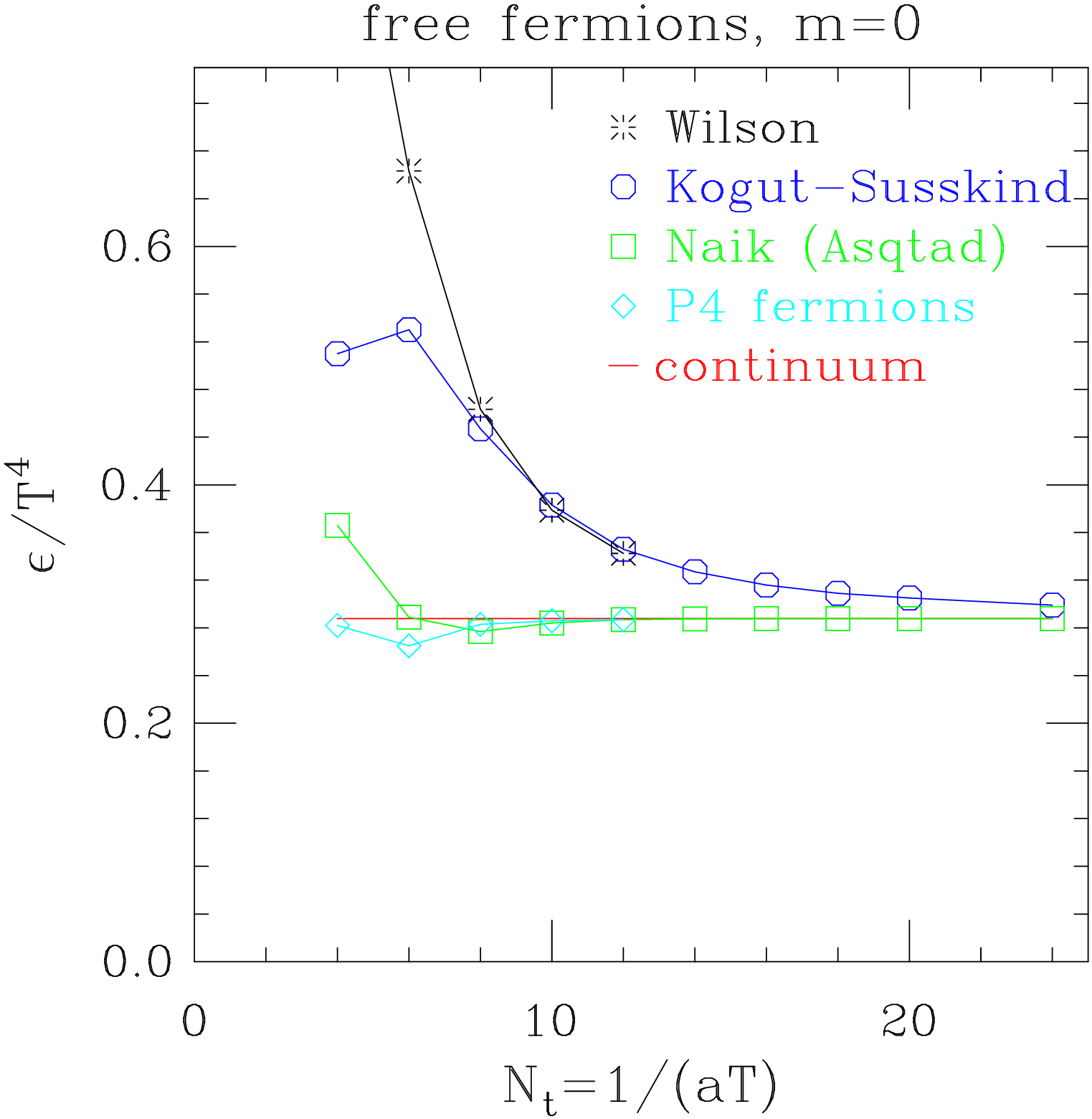}
&
\includegraphics[width=2.5in]{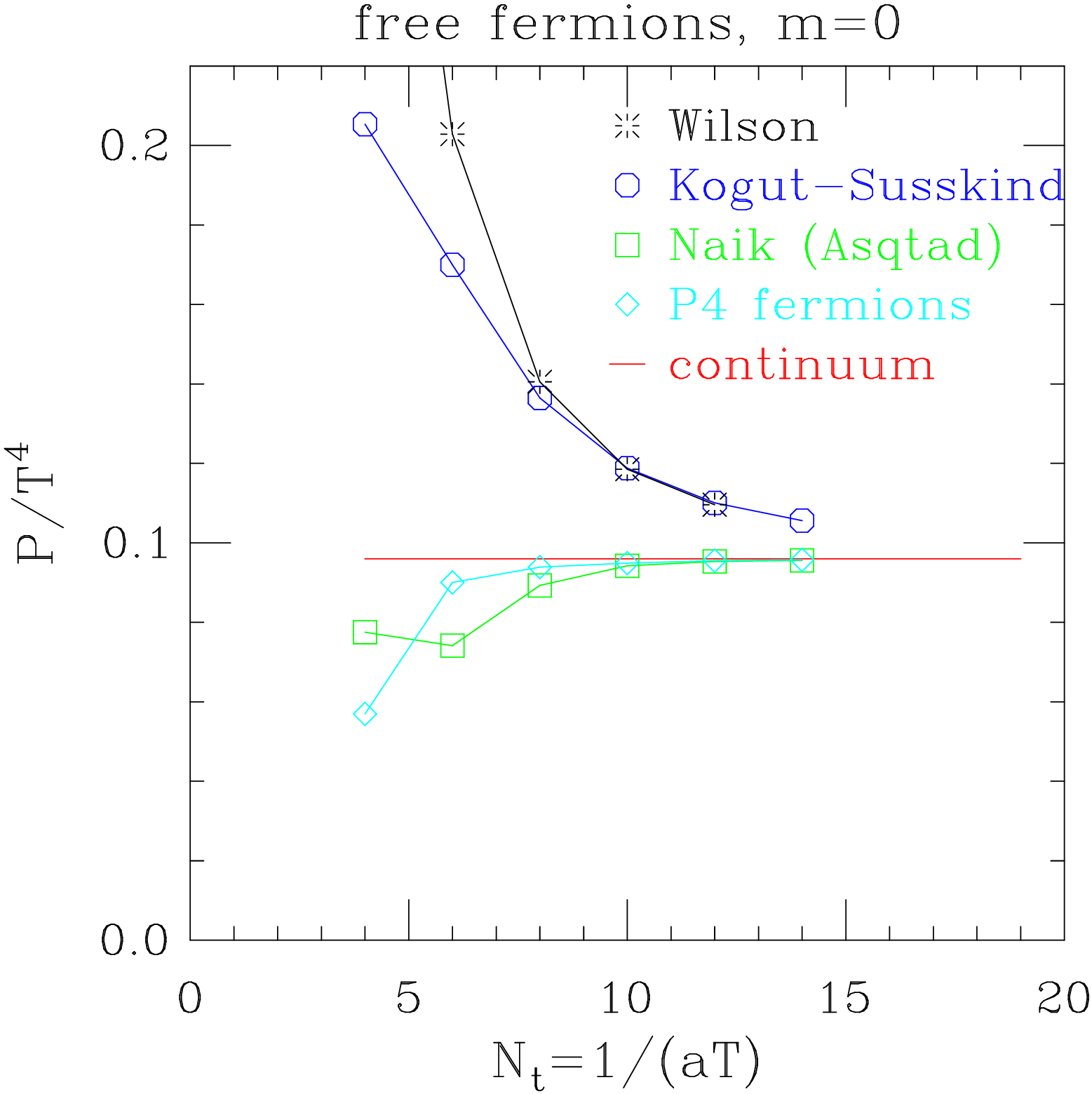}
\end{tabular}
\end{center}
\vspace*{-9mm}
\caption{The energy density (left) and pressure (right) of free massless
fermions as function of temporal lattice size $N_t$. For free fermions
the ``Asqtad'' action reduces to the Naik action. The ``p4'' action is the
improved staggered fermion action preferred by the Bielefeld group
\protect\cite{p4_fT}.
\label{fig:free_e_p} 
}
%\vspace*{-10mm}
\end{figure}

The three-link ``Naik'' term insures a good dispersion relation and thereby
helps decrease the lattice artifacts in energy density and pressure at
small temporal lattice size $N_t$, as can be seen in Fig.~\ref{fig:free_e_p}.

Based on the zero temperature simulations of Ref.~\cite{MILC_Spec} and the
estimate of $T_c \sim 150 - 170$ MeV \cite{FK_Schl01} we deduce that for
$N_t=8 - 10$ the kaon will be heavier
than the heaviest non-Goldstone pion at the finite temperature transition.

\begin{figure}
\begin{center}
\begin{tabular}{c c}
\includegraphics[width=2.5in]{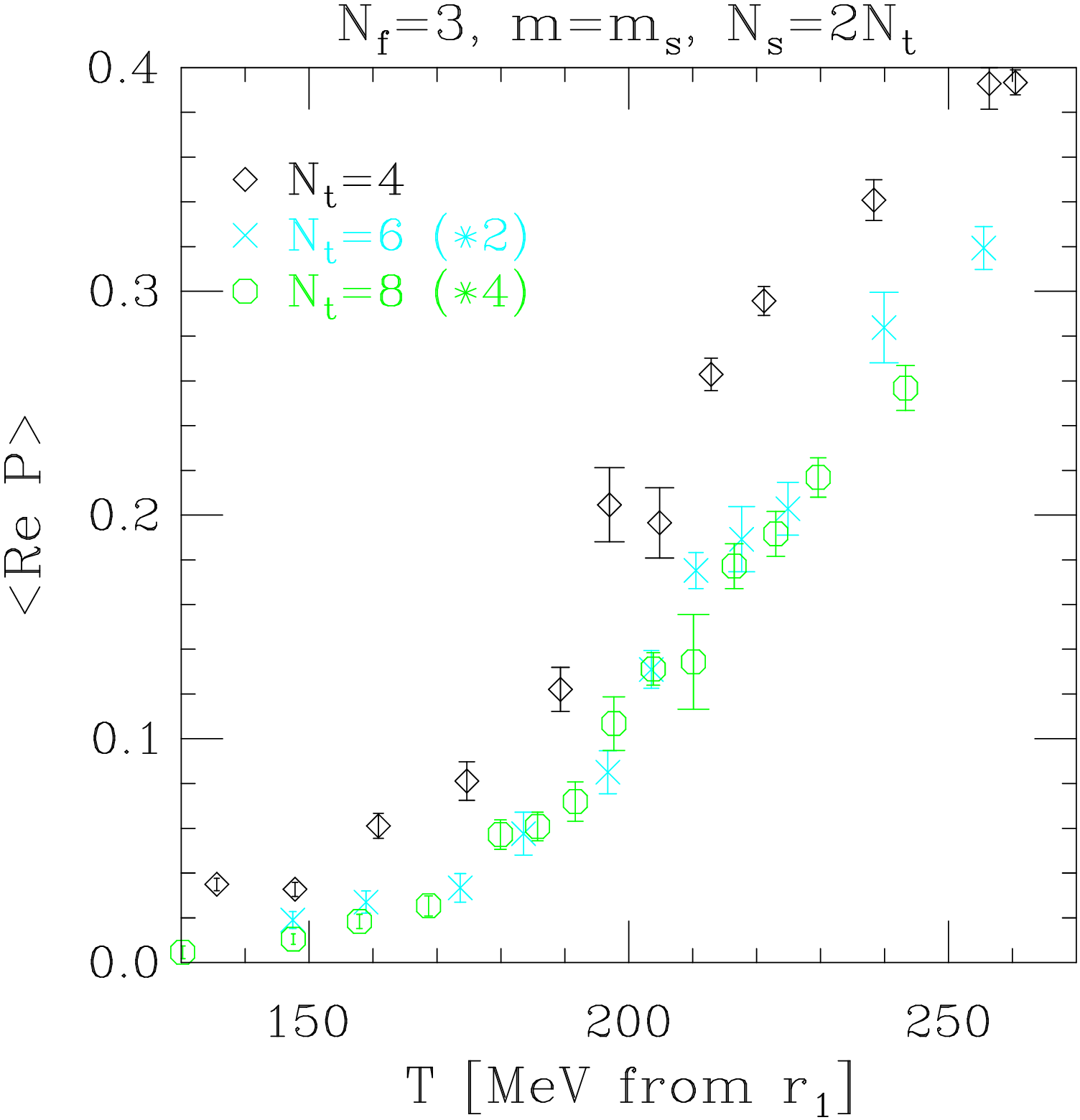}
&
\includegraphics[width=2.5in]{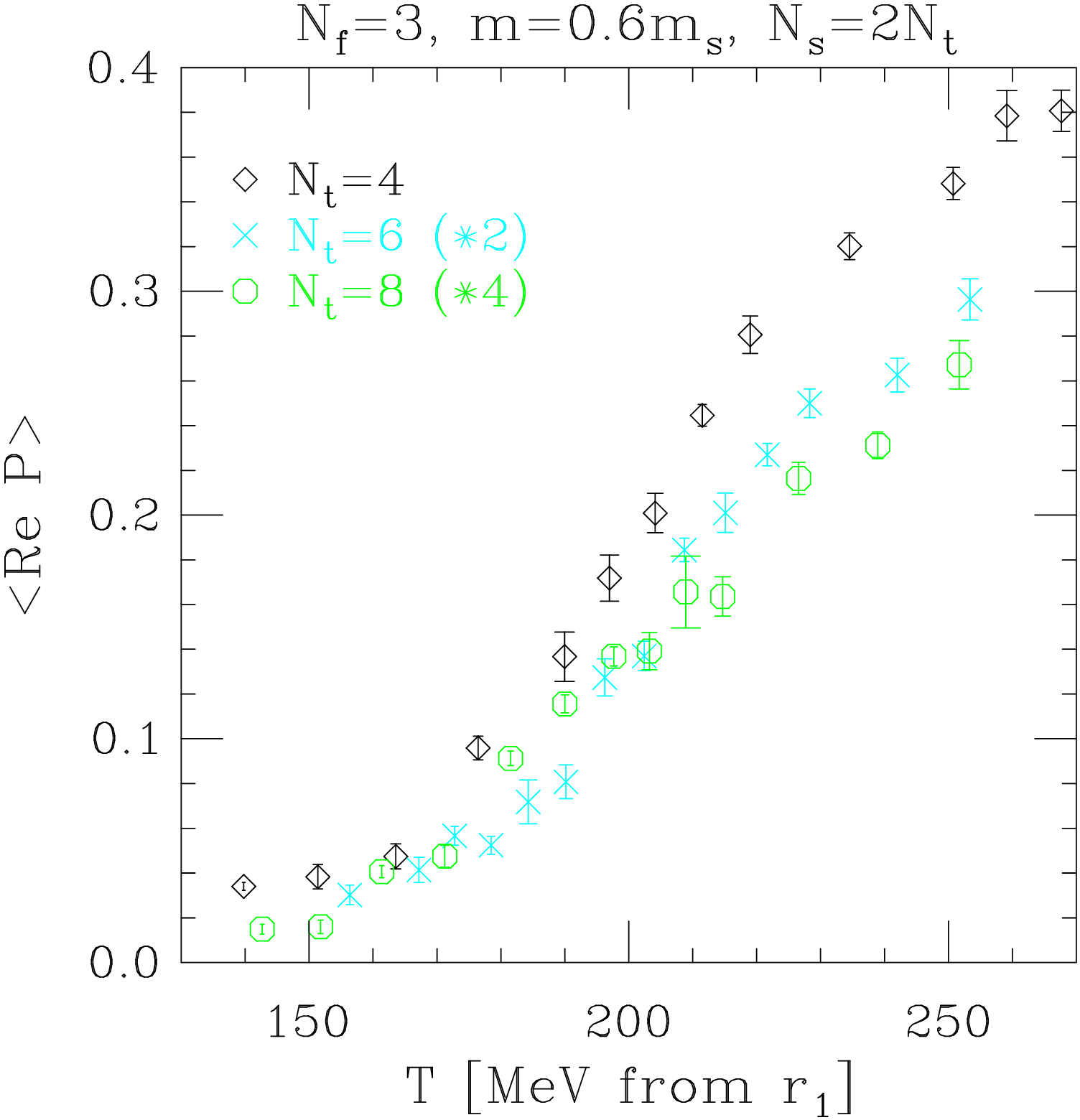}
\end{tabular}
\end{center}
\vspace*{-9mm}
\caption{Real part of the Polyakov line for three flavors with $m_q \approx m_s$
(left) and $m_q \approx 0.6 m_s$ (right). The data for $N_t=6$ have been
multiplied by two and the data for $N_t=8$ by four.
The spatial lattice sizes are $N_s = 2N_t$.
\label{fig:polr_m05_m03} 
}
\vspace*{3mm}
%\vspace*{-10mm}
%\end{figure}
%
%\begin{figure}
\begin{center}
\begin{tabular}{c c}
\includegraphics[width=2.5in]{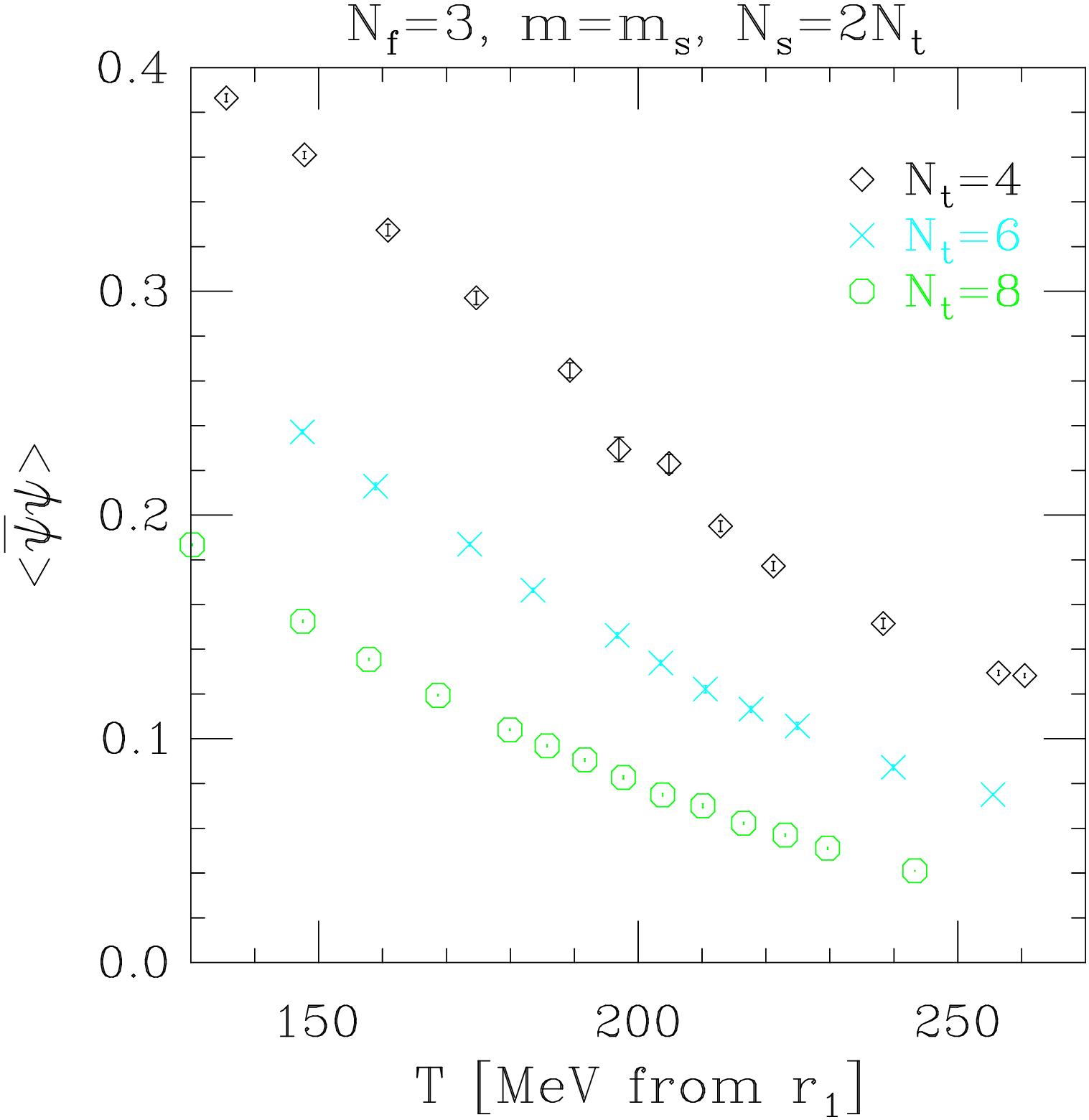}
&
\includegraphics[width=2.5in]{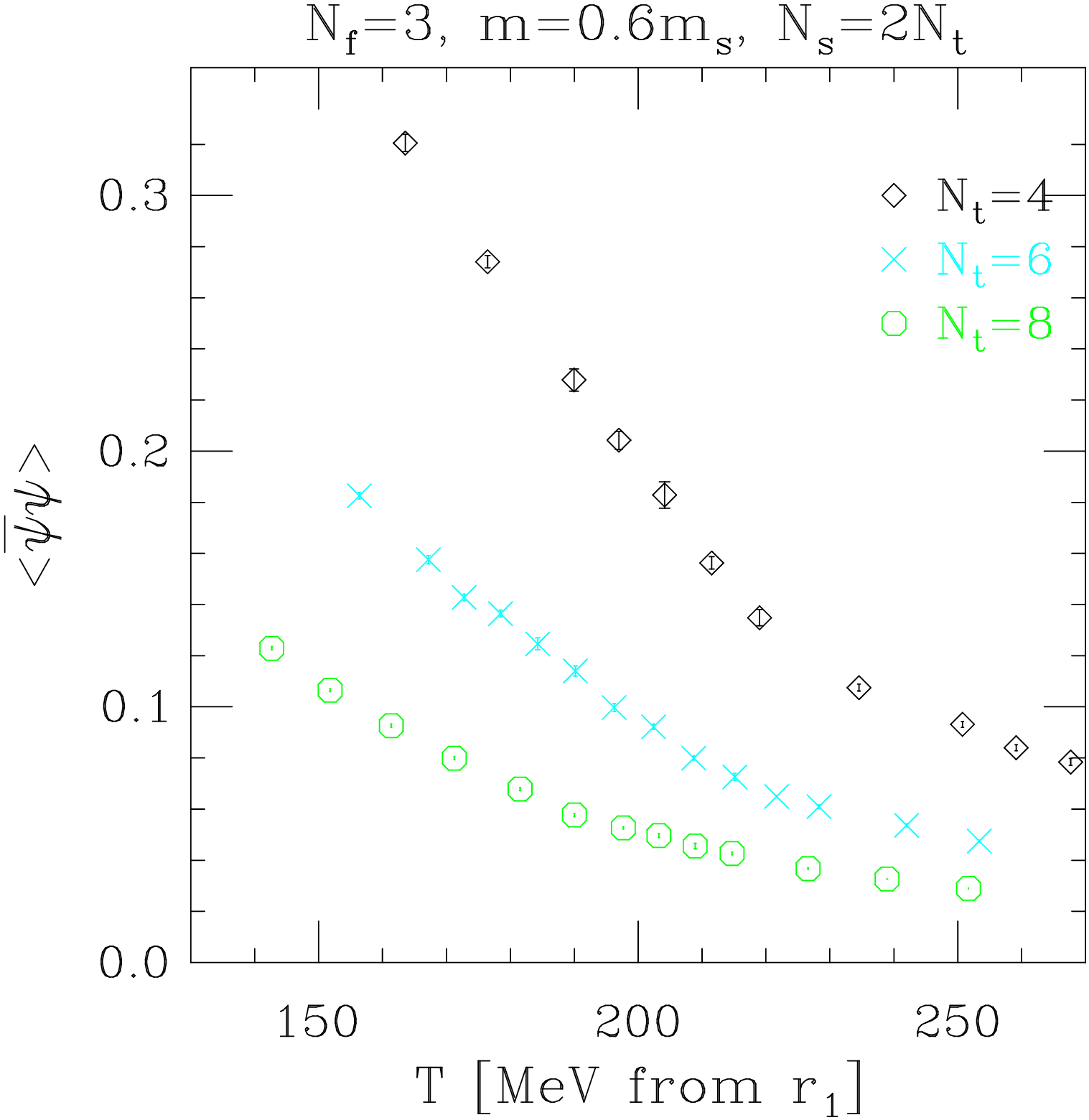}
\end{tabular}
\end{center}
\vspace*{-9mm}
\caption{$\< \bar \psi \psi \>$ for three flavors with $m_q \approx m_s$ (left)
and $m_q \approx 0.6 m_s$ (right).
\label{fig:pbp_m05_m03} 
}
%\vspace*{3mm}
%\vspace*{-10mm}
\end{figure}

We have zero temperature results, in particular, the value of the (bare) strange
quark mass, at fixed lattice spacing $a \sim 0.13$ fm and $a \sim 0.2$ fm.
Since we want to keep the physical quark masses approximately constant,
when we vary the temperature which, at fixed $N_t=1/(aT)$, means varying the
gauge coupling $\beta$, we interpolated (extrapolated) between the values
at the two lattice spacings.

Given a gauge coupling $\beta$ and quark mass $am_q$, for the three-flavor
simulations, or quark masses $am_s$ and $am_{u,d}$, for $2+1$ flavors, we
determined the tadpole factor $u_0 = (\Tr U_p/3)^{1/4}$ self-consistently
in short zero temperature simulations on $L^4$ lattices.

To convert lattice scales to physical scales we interpolated $a/r_1$ with
the form advocated by Allton \cite{Allton}
\begin{eqnarray}
%a/r_1 &=& c_0 f(g^2_0) \left[ 1 + c_2 g^2_0 f^2(g^2_0) \right] \nonumber \\
%f(g^2_0) &=& (b_0 g^2_0)^{-b_1/(2b^2_0)} \exp \left( - \frac{1}{2b_0 g^2_0}
a/r_1 = c_0 f(g^2_0) \left[ 1 + c_2 g^2_0 f^2(g^2_0) \right] ~, \qquad \qquad
f(g^2_0) = (b_0 g^2_0)^{-b_1/(2b^2_0)} \exp \left( - \frac{1}{2b_0 g^2_0}
 \right) ~.
\end{eqnarray}
For the (one-loop, tadpole) Symanzik improved gauge action the bare coupling
$g^2_0$ is related to $\beta$ by $g^2_0 = 10/\beta$. $b_{0,1}$ are the
universal first two coefficients of the beta-function for three massless
flavors, and $c_{0,2}$ are determined from the measured values of $a/r_1$
at $a \sim 0.13$ fm and $a \sim 0.2$ fm.

The Hybrid Molecular Dynamics R-algorithm was used for all simulations,
with two noise vectors for the $2+1$ flavor simulations.

\section{RESULTS FOR THREE DEGENERATE FLAVORS}

For our first simulation with $m_q \approx m_s$ we used a linear interpolation
of $am_q(\beta)$ in $\beta$. Since the range of interpolation/extrapolation
is quite large, we subsequently used a linear interpolation of
$\log(am_q(\beta))$, compatible with the leading behavior of $a$ (with
$m_q$ fixed) as function of $\beta$. In the region of the finite temperature
transition or crossover for lattices with temporal extent $N_t=6$ and 8,
the difference between the two interpolation schemes is small enough that
it will not affect our results significantly.

We show in Fig.~\ref{fig:polr_m05_m03} the real part of the Polyakov line
and in Fig.~\ref{fig:pbp_m05_m03} the condensate as function of the
temperature, with the physical scale obtained via $r_1$, for our simulations
with three degenerate quarks of mass $m_q \approx m_s$ and
$m_q \approx 0.6 m_s$. The Polyakov line shows a crossover from confined
behavior at low temperature to a deconfined behavior at high temperature.
The condensate decreases as the temperature is increased, but shows no sign
of a transition or sharp crossover.

\begin{figure}
\begin{center}
\begin{tabular}{c c}
\includegraphics[width=2.5in]{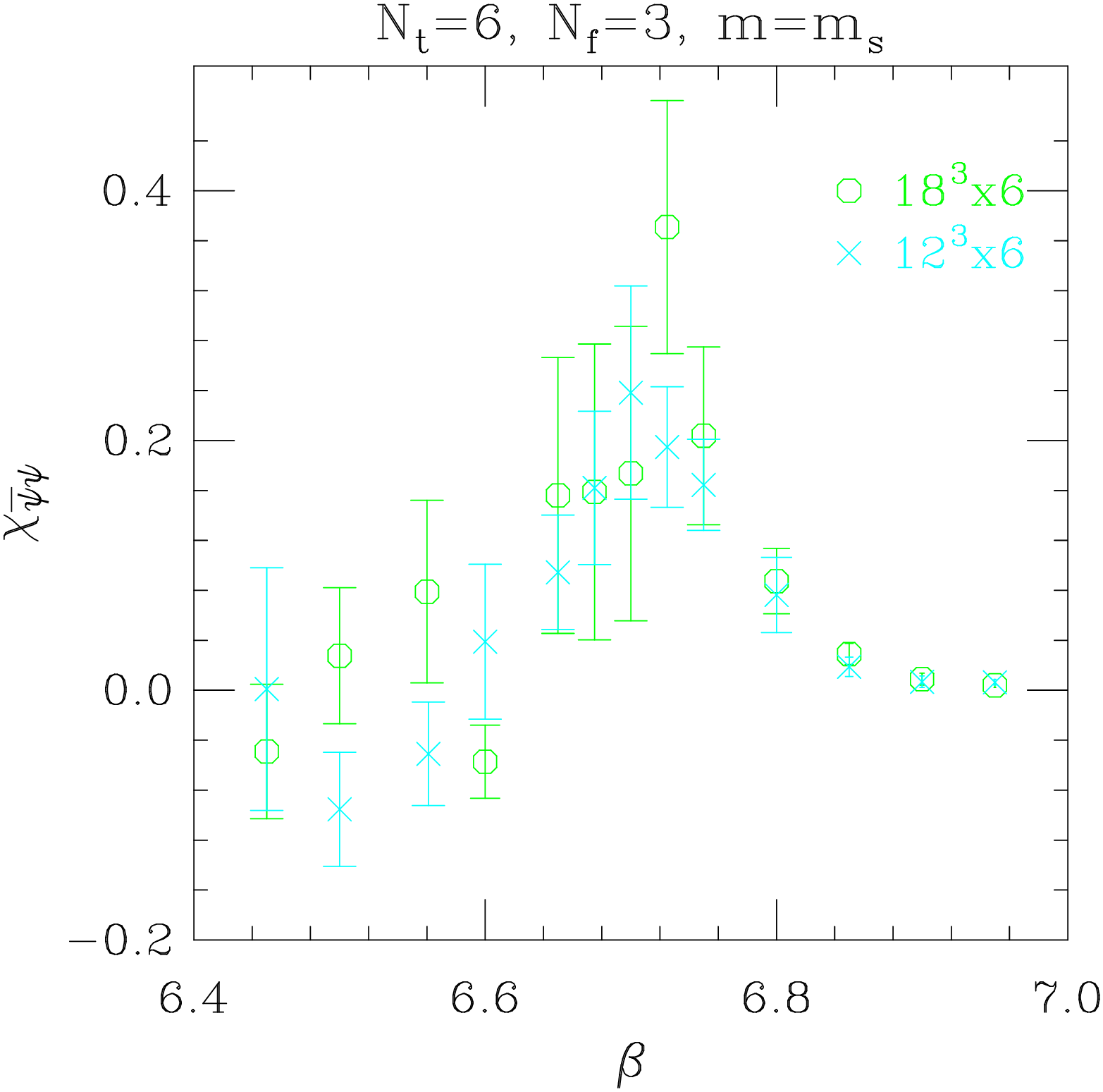}
&
\includegraphics[width=2.5in]{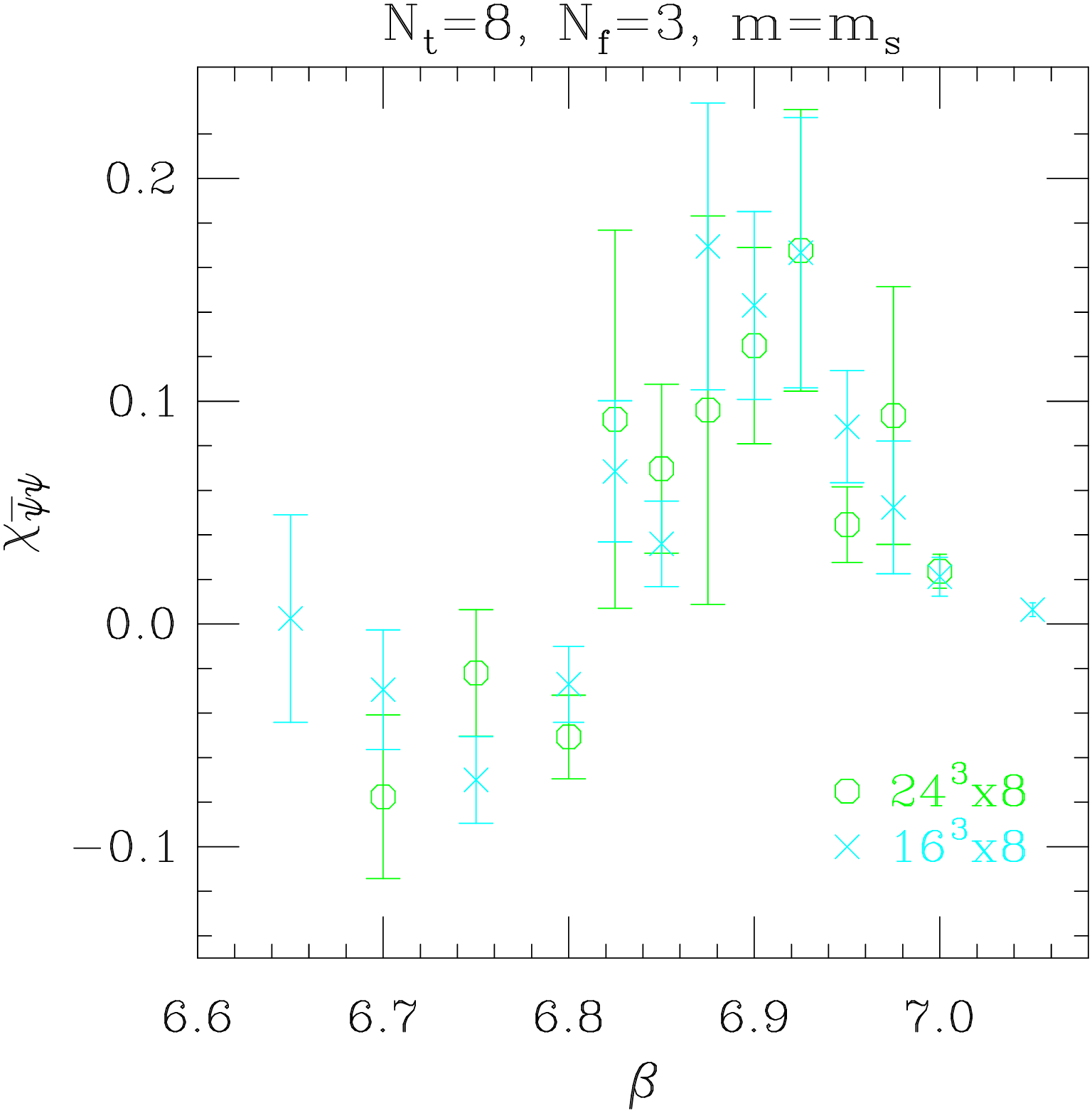}
\end{tabular}
\end{center}
\vspace*{-9mm}
\caption{The $\bar\psi \psi$ susceptibility for three flavors with
$m_q \approx m_s$ for lattices with $N_t=6$ (left) and $N_t=8$ (right).
\label{fig:chi_t6_t8_m05} 
}
%\vspace*{-10mm}
\end{figure}

In Fig.~\ref{fig:chi_t6_t8_m05} we show the chiral susceptibility,
$\chi_{\bar\psi \psi} = \partial \< \bar\psi \psi \> / \partial m_q$ from
the simulations with $m_q \approx m_s$ for lattices with $N_t=6$ and 8.
A peak is seen in both cases. But the peak height does not increase when
we change the spatial lattice size from $N_s = 2N_t$ to $3N_t$, indicating
that we are observing a smooth crossover from a chirally symmetric phase at
high temperature (large $\beta$) to a phase of broken chiral symmetry at low
temperature. The location of the peak coincides approximately with the
steep increase of the Polyakov line in Fig.~\ref{fig:polr_m05_m03}. The
chiral susceptibility for $m_q \approx 0.6 m_s$ looks similar to
Fig.~\ref{fig:chi_t6_t8_m05}, but in that case we have, so far, results
only for one spatial size, $N_s = 2N_t$.

\section{RESULTS FOR 2+1 FLAVORS}

\begin{figure}
\begin{center}
\begin{tabular}{c c}
\includegraphics[width=2.5in]{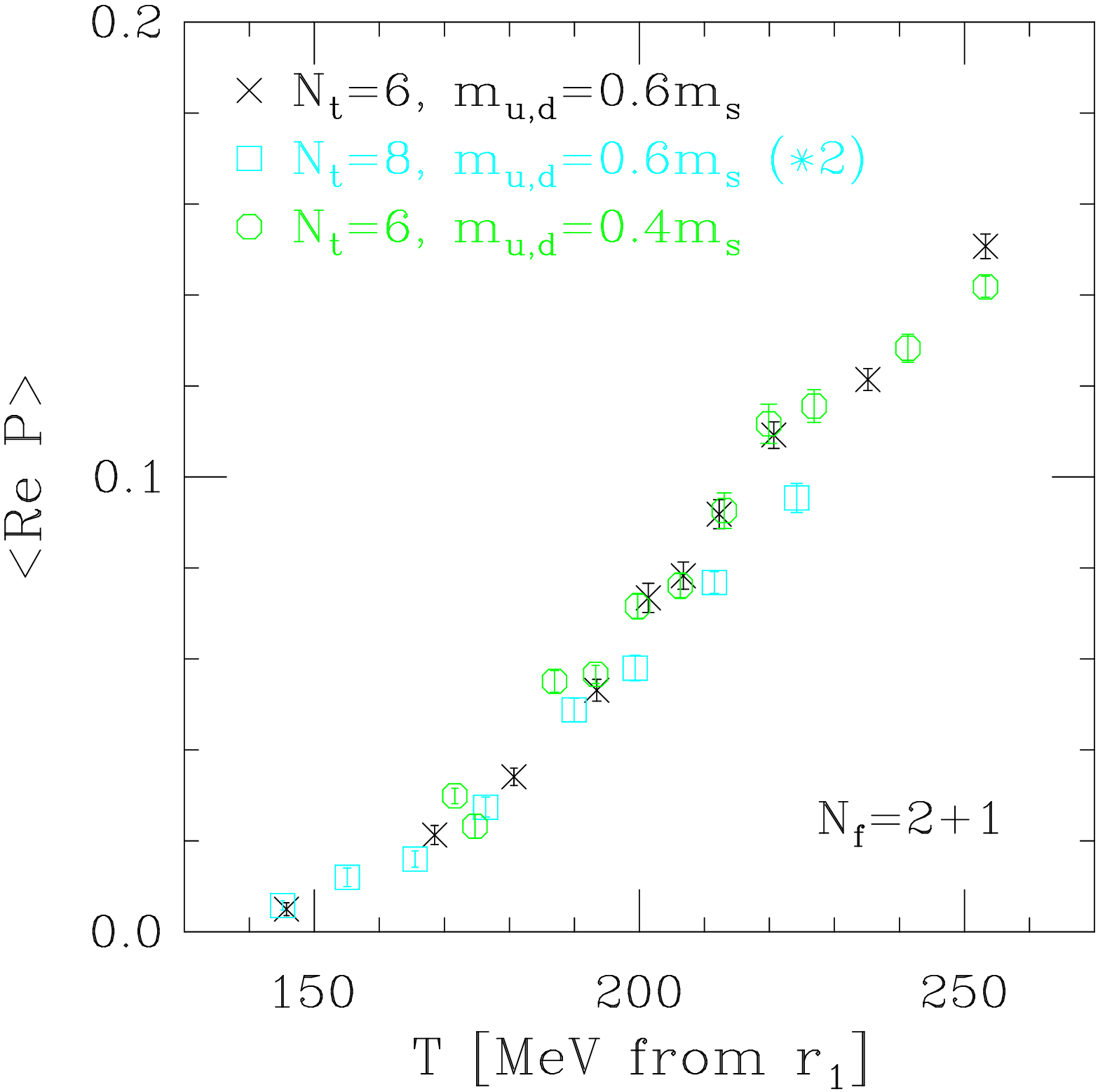}
&
\includegraphics[width=2.5in]{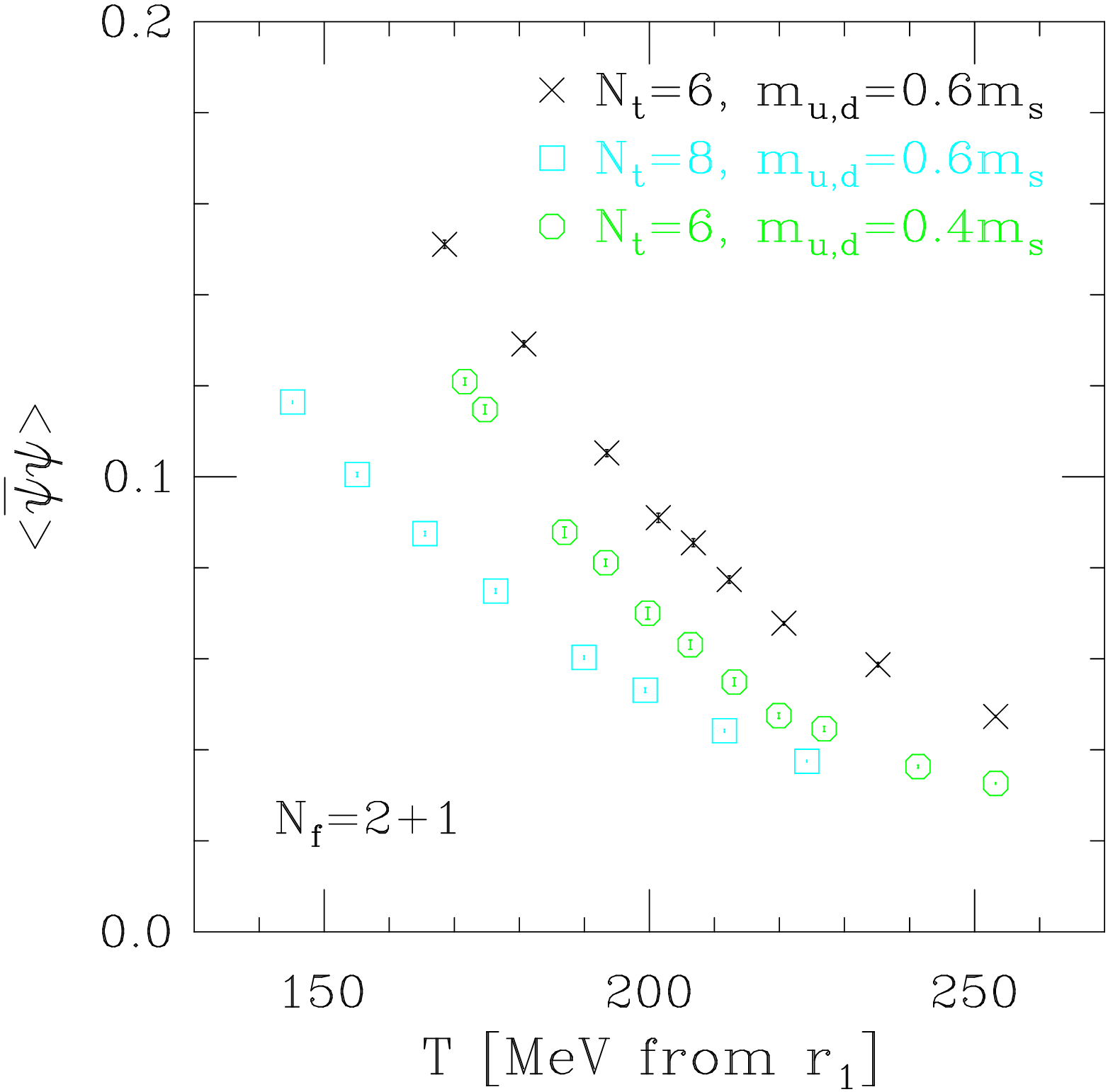}
\end{tabular}
\end{center}
\vspace*{-9mm}
\caption{Real part of the Polyakov line (left) and $\< \bar \psi \psi \>$ for
the light quarks (right) for 2+1 flavors. The Polyakov line data for $N_t=8$
has been multiplied by two.
\label{fig:polr_pbp_f21} 
}
%\vspace*{-10mm}
\end{figure}

We have also preliminary results from simulations with 2+1 dynamical flavors,
with the heavier quark kept at the strange quark mass, and the lighter
masses at $m_{u,d} = 0.6 m_s$ and $0.4 m_s$. The real part of the Polyakov
line and the light quark condensate from these simulations are shown in
Fig.~\ref{fig:polr_pbp_f21}. Again we see a crossover behavior, particularly
for the Polyakov line. For these simulations we do not have the data yet
to compute the chiral susceptibility.

\section{CONCLUSIONS}

We have made first simulations to explore the finite temperature phase diagram
with an improved staggered fermion action, the ``Asqtad'' action, which reduces
flavor symmetry breaking so that all pions are lighter than the kaon already
at larger lattice spacing, and which improves rotational symmetry and the
dispersion relation, leading to diminished lattice artifacts in energy
density and pressure.

For three flavor simulations down to quark mass $m_q \approx 0.6 m_s$,
and for 2+1 flavor simulations with the light quark masses down to
$m_{u,d} = 0.4 m_s$ while the heavier quark is kept at the strange quark
mass, we observed finite temperature crossover behavior, but so far no
sign of a genuine phase transition. This result is compatible with other
recent simulations \cite{JLQCD,crit_point} which found phase transitions
only at quark masses lighter than those studied by us so far. The temperature
at the crossover, with the scale set by $r_1$, $T_c \sim 190$ -- 200 MeV,
is a little higher than expected from previous determinations
\cite{p4_fT,FK_Schl01}. This is presumably also due to our quark
masses still being rather high.

This work is supported by the US National Science Foundation and
Department of Energy and used computer resources at Florida State
University (SP), NERSC, NPACI, FNAL, and the University of Utah (CHPC).

\end{document}